\documentclass[prl,twocolumn,aps,superscriptaddress,showpacs]{revtex4}

\usepackage{amsfonts}
\usepackage{mathrsfs}
\usepackage{amsmath}
\usepackage{color}
\usepackage{graphicx}
\usepackage{bm}
\usepackage{amssymb}
\usepackage{xspace}
\usepackage{epstopdf}
\usepackage{dcolumn}
\usepackage{longtable}
\usepackage{multirow}
\usepackage[colorlinks=true, letterpaper=true, pdfstartview=FitV, linkcolor=blue, citecolor=blue, urlcolor=blue]{hyperref}

\begin{document}

\title{$d$-Orbital Topological Insulator and Semimetal in Antifluorite Cu$_2$S Family: Contrasting Spin Helicities, Nodal Box, and Hybrid Surface States}

\author{Xian-Lei Sheng}
\affiliation{Research Laboratory for Quantum Materials, Singapore University of Technology and Design, Singapore 487372, Singapore}
\affiliation{Department of Applied Physics, Key Laboratory of Micro-nano Measurement-Manipulation and Physics (Ministry of Education), Beihang University, Beijing 100191, China}

\author{Zhi-Ming Yu}
\affiliation{Research Laboratory for Quantum Materials, Singapore University of Technology and Design, Singapore 487372, Singapore}

\author{Rui Yu}
\affiliation{School of Physics and Technology, Wuhan University, Wuhan 430072, China}

\author{Hongming Weng}
\email{hmweng@iphy.ac.cn}
\affiliation{Beijing National Laboratory for Condensed Matter Physics, and Institute of Physics, Chinese Academy of Sciences, Beijing 100190, China}
\affiliation{Collaborative Innovation Center of Quantum Matter, Beijing, China}

\author{Shengyuan A. Yang}
\email{shengyuan\_yang@sutd.edu.sg}
\affiliation{Research Laboratory for Quantum Materials, Singapore University of Technology and Design, Singapore 487372, Singapore}


\begin{abstract}
We reveal a class of three-dimensional $d$-orbital topological materials in the antifluorite Cu$_2$S family. Derived from the unique properties of low-energy $t_{2g}$ states, their phases are solely determined by the sign of spin-orbit coupling (SOC): topological insulator for negative SOC, whereas topological semimetal for positive SOC; both having Dirac-cone surface states but with contrasting helicities. With broken inversion symmetry, the semimetal becomes one with a nodal box consisting of butterfly-shaped nodal lines that are robust against SOC. Further breaking the tetrahedral symmetry by strain leads to an ideal Weyl semimetal with four pairs of Weyl points. Interestingly, the Fermi arcs coexist with a surface Dirac cone on the (010) surface, as required by a $Z_2$-invariant.
\end{abstract}

\maketitle



When individual atoms are brought together to form crystalline solids, the atomic orbitals overlap and form extended Bloch states. In the energy-momentum space, discrete atomic levels evolve into dispersive electronic bands. The interaction between orbitals and with further coupling to spin may generate inverted band ordering and lead to topological states of matter, which is a focus of recent physics research~\cite{RevModPhys.82.3045,RevModPhys.83.1057,RevModPhys.88.021004}.
It is now established that nontrivial topology can occur for both gapped (insultor) and gapless (semimetal) systems. For topological insulators (TIs), an invariant is defined for the bulk valence bands below the gap~\cite{RevModPhys.82.3045}; whereas for topological semimetals (TSMs), the characterization is on the topology of band-crossings near the Fermi level~\cite{Volovik2003,Zhao2013c}, leading to a variety of TSMs, among which the Weyl semimetal and nodal-line semimetal states with respective 0D and 1D band-crossings are attracting great interest and actively searched for~\cite{WanXG_Weyl,Murakami2007,WengHM_2015TaAs,Huang2015,Lv2015,Xu613,Xu2015a,Yang2015,PhysRevB.92.045108,PhysRevLett.115.036806,PhysRevLett.115.036807,Chen2015,PhysRevB.93.241202,PhysRevB.94.165201,PhysRevX.6.031003}. The nontrivial bulk topology manifests on the sample surface as the existence of protected surface states: TIs have Dirac-cone like surface states with spin-momentum-locking~\cite{RevModPhys.82.3045,RevModPhys.83.1057}; whereas Weyl semimetals possess open Fermi arcs connecting pairs of projected Weyl points on the surface~\cite{WanXG_Weyl}.

In forming topological band structures, the orbital character of bands plays an important role, e.g., it determines the band inversion, the low-energy quasiparticle dispersion, and furthermore, the type and strength of the effective spin-orbit coupling (SOC). For almost all the TIs identified so far, the low-energy bands are of $s$- and/or $p$-orbital character~\cite{YaoYG2012_TIreview,Ando2013}. Meanwhile, it is known that $d$-orbitals could host interesting SOC physics. For instance, with tetrahedral coordination, the five $d$-orbitals will split into an $e_g$ doublet and a $t_{2g}$ triplet. With SOC, the $t_{2g}$ states can exhibit a unique property that its effective SOC is negative, with $j=1/2$ doublet energetically higher than the $j=3/2$ quartet. This mechanism has inspired works in exploring TIs with negative SOC. However, in the few examples predicted to date~\cite{Vidal:2012cr,Sheng_TlN,Virot:2011fn}, the low-energy bands are still dominated by $s$ and $p$ characters, while the $d$-bands are away from the Fermi level and only act indirectly. Thus, one may wonder whether we can find a genuine $d$-orbital TI, and how would the $d$ character produce any new physics?

\begin{figure}[tbp]
\centerline{\includegraphics[clip,width=9cm]{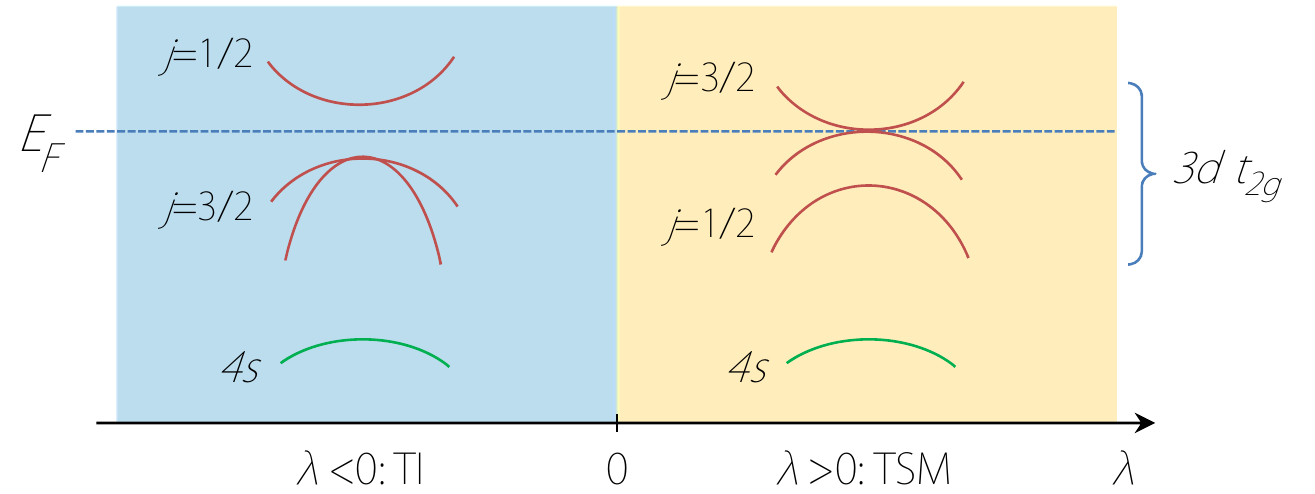}}
\caption{Schematic of the essential band features of Cu$_2$S-family materials. The low-energy bands are dominated by 3$d$ $t_{2g}$-orbitals, which are splitted by SOC into $j=1/2$ and $j=3/2$ states (each line here is assumed doubly spin degenerate). The sign of SOC $\lambda$ determines the phase: TI for negative SOC, while TSM for positive SOC. The band inversion is between the $t_{2g}$ band and the $4s$ band.}
\label{fig1}
\end{figure}

Here, we answer the above questions by revealing intriguing $d$-orbital topological phases in the Cu$_2$S material family with antifluorite structure. By first-principles calculations, we show that the low-energy bands in these materials are dominated by the $t_{2g}$-orbitals, of which the sign of effective SOC can be made either negative or positive by tuning the orbital interaction. Remarkably, this sign completely determines the phase. As illustrated in Fig.~\ref{fig1}, enforced by band filling, the system must be an insulator (semimetal) when the sign of SOC is negative (positive). Moreover, in both cases, there is band inversion between the cation $s$-band and the $t_{2g}$-band, hence both phases are topological. We explicitly show that the Dirac-cone surface states in the two phases have opposite helicities in spin-momentum-locking, consistent with their sign of SOC. Furthremore, novel features are observed for the TSM phase. In ternary compounds with intrinsic inversion asymmetry, the system becomes a novel semimetal with a nodal box comprising butterfly-shaped nodal lines. In known examples of nodal-line materials, the nodal lines are unstable when SOC is included; in contrast, the butterfly nodal line here is stabilized with SOC. With further lowering of cubic symmetry by strain, an ideal Weyl semimetal emerges with four pairs of Weyl points lying exactly at the Fermi level. We find the phenomenon of hybrid surface states, i.e., coexisting Fermi-arc and Dirac-cone surface states on certain surfaces, as required by a bulk $Z_2$ invariant. Our predicted features in the bulk band structure and the surface states (including the spin texture) can be readily probed by the ARPES experiment.

\begin{figure}[tbp]
\centerline{\includegraphics[clip,width=9.2cm]{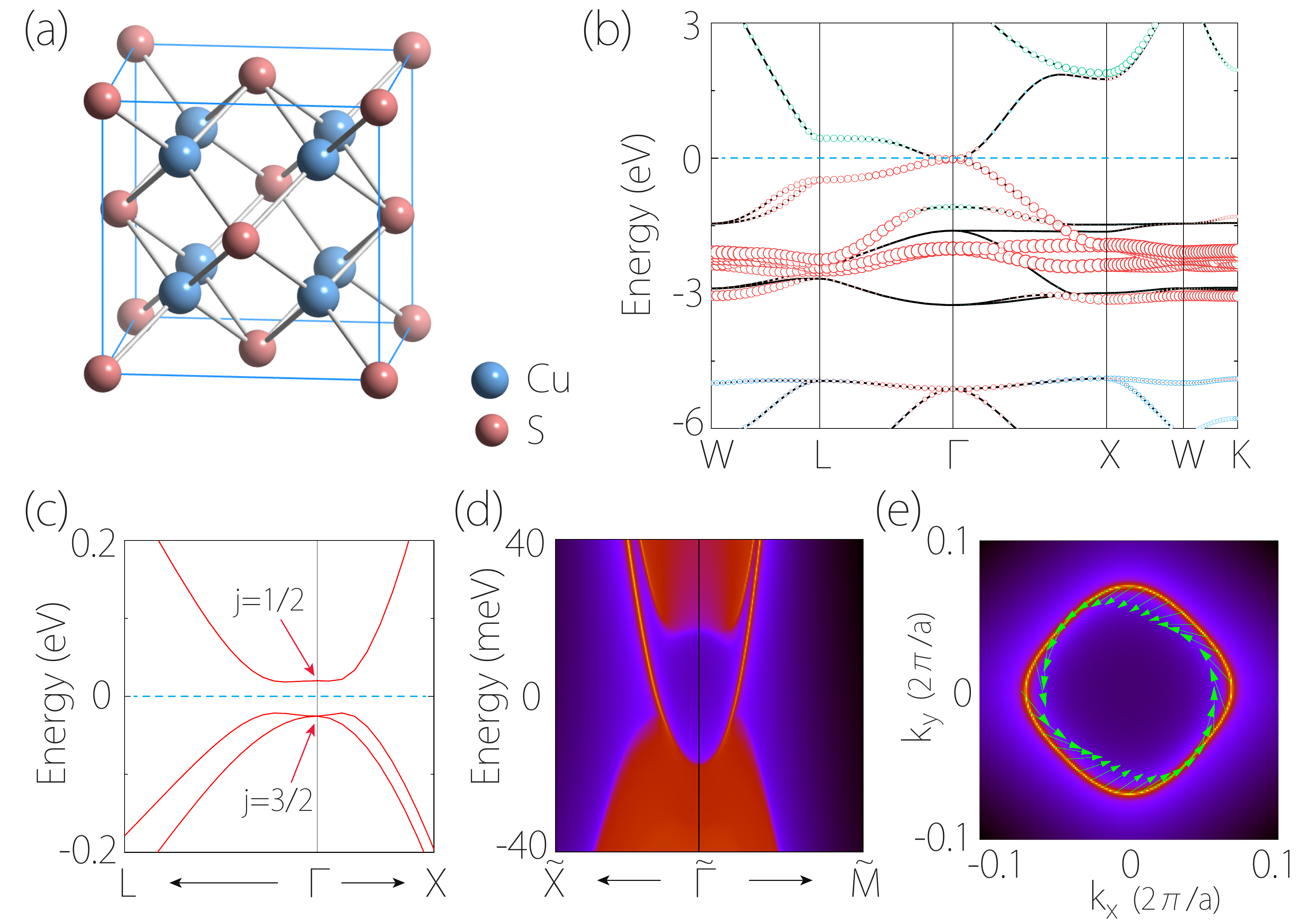}}
\caption{(a) Conventional unit cell of Cu$_2$S lattice. (b) Band structure of Cu$_2$S. The colored circles indicate the weight of Cu-$t_{2g}$ (red), Cu-$4s$ (green), S-$3p$ (blue) orbital characters. (c) Enlarged view of (b) near the Fermi level, showing the splitting between $j=1/2$ and $3/2$ states due to negative SOC. (d) Projected spectrum on (001) surface, and (e) the corresponding Fermi circle, exhibiting a right-handed spin-momentum-locking (spin-polarization marked by the green arrows). }
\label{fig2}
\end{figure}

The Cu$_2$S-family materials typically have three structures (denoted as $\alpha$, $\beta$, and $\gamma$). Here, we focus on the $\alpha$-phase structure~\cite{Cu2S_5.725,Cu2Se_constant,CuAgS_constant}, also known as the antifluorite structure, having the space group $Fm\bar{3}m$ (No.~225). The conventional unit cell has a cubic shape and the structure can be regarded as a double nested zinc-blende lattice. As shown in Fig.~\ref{fig2}(a), for binary compounds like Cu$_2$S and Cu$_2$Se, an inversion center is preserved. However, inversion symmetry is broken for ternary compounds like CuAgSe (see Fig.~\ref{fig4}(a)) where the Cu sites in one of the nested zinc-blende lattices are occupied by Ag. We perform first-principles calculations based on the density functional theory (DFT). The calculation details and the structural parameters are in the Supplemental Material~\cite{SI}.
In the following discussion, we shall mainly focus on the representatives Cu$_2$S, Cu$_2$Se, and CuAgSe.

{\color{blue}{\em Negative SOC: TI.}}---In Cu$_2$S, each Cu is surrounded by a tetrahedron of S atoms. The tetrahedral crystal field splits the Cu 3$d$-orbitals into $e_g$ and $t_{2g}$ states, with $t_{2g}$ having a higher energy. Focusing on the $t_{2g}$ states, the interaction between the two Cu atoms in a primitive cell leads to bonding and antibonding states, with the latter energetically higher than the former. In Cu$_2$S, the Cu $d$-orbitals have higher energy than that of S-3$p$ orbitals, so that the $p$-$d$ hybridization further pushes the $t_{2g}$ antibonding states up to the Fermi level. The $t_{2g}$ triplet (containing $d_{xy}$, $d_{xz}$, and $d_{yz}$ orbitals) have an effective orbital moment $\ell=1$, which are then split by the SOC $\lambda \bm \ell\cdot \bm s$ into a $j=1/2$ doublet and a $j=3/2$ quartet. As mentioned, a unique feature for $t_{2g}$ is that the SOC can be negative with $\lambda<0$, opposite to the SOC splitting of $p$-orbitals~\cite{Cardona1963,Shindo1965}.

Indeed, as shown in Fig.~\ref{fig2}, our DFT result confirms the above picture. The low-energy bands near the Fermi level are mainly from the $t_{2g}$ states, and the S-3$p$ bands are below $-4$ eV. Due to combined inversion symmetry and time reversal symmetry, each band is spin degenerate. Around $\Gamma$-point, one observes that: (i) the originally degenerate $t_{2g}$ states are split by SOC, and the $j=1/2$ doublet is higher than the $j=3/2$ quartet in energy, showing a negative SOC; (ii) the Cu-4$s$ states dive below the $t_{2g}$ states by about 1.1 eV, indicating an inverted band ordering. Band filling dictates that the Fermi level lies exactly in the gap ($\sim 55$ meV) between the $j=1/2$ and $j=3/2$ states (Fig.~\ref{fig2}(c)). The band inversion at a single time reversal invariant momentum (TRIM) point directly indicates that the system is a strong TI. To further confirm the nontrivial band topology, we calculate the $Z_2$ invariant for the bulk band structure. With inversion symmetry, the task is simplified by analyzing the product of parity eigenvalues at the eight TRIM points~\cite{Fu2007}. We find that this product is positive at $\Gamma$ and negative at other TRIMs, in accordance with our analysis of the band inversion, which leads to a strong TI with $Z_2$ indices $(1;000)$.

The hallmark of TI is the existence of protected Dirac-cone surface states with spin-momentum-locking. In Fig.~\ref{fig2}(d), we plot the calculated surface energy spectra for (001) surface, clearly showing a single surface Dirac-cone. Here the Dirac point is buried in the bulk valence bands. Notice that for a constant energy above the Dirac point, the surface states show a right-handed spin-momentum-locking pattern (see Fig.~\ref{fig2}(e)). This is consistent with the negative SOC, and is in contrast with almost all other TIs (which have left-handed helicity due to positive SOC)~\cite{ZhangW_NJP,PhysRevLett.111.066801}. The same feature is also observed for other surfaces. All these evidences confirm that Cu$_2$S is a $d$-orbital TI.

\begin{figure}[tbp]
\centerline{\includegraphics[clip,width=9.2cm]{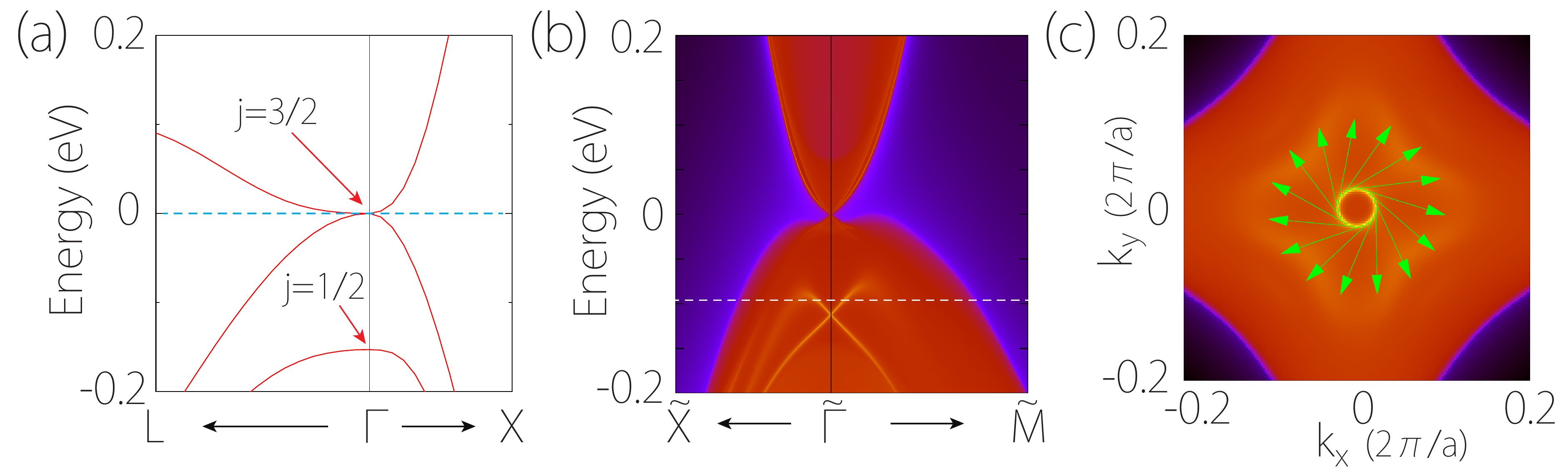}}
\caption{Results for Cu$_2$Se. (a) Low-energy band structure. (b) Projected spectrum on (001) surface showing Dirac-cone surface states buried in the bulk valence band. (c) Surface states at $-0.1$ eV (marked in (b)), exhibiting a left-handed spin-momentum-locking.}
\label{fig3}
\end{figure}

{\color{blue}{\em Positive SOC: TSM.}}---The effective SOC strength $\lambda$ could be tuned by varying the crystal environment, especially through the interaction with the anion $p$-orbitals. The effect depends on both the interaction strength and the SOC of $p$-orbitals. Consider Cu$_2$Se which has the same structure as Cu$_2$S. Compared with S-3$p$, the Se-4$p$ orbitals are closer to the Cu-$t_{2g}$ orbitals in energy, and they also possess a stronger SOC (which is positive). Consequently, one expects that the interaction with Se-4$p$ would decrease the negativeness of $\lambda$ for Cu-$t_{2g}$ states.

This speculation is confirmed by our DFT result. We find that the composition of the low-energy bands in Cu$_2$Se is similar to that of Cu$_2$S, however, its $j=3/2$ quartet is above the $j=1/2$ doublet (see Fig.~\ref{fig3}(a)), indicating that SOC has changed from negative to positive. Band filling dictates that the quartet is half-filled, so that the Fermi level intersects with the degenerate states and the system must be a semimetal.

Note that the band inversion near $\Gamma$-point between the Cu-4$s$ states and the $t_{2g}$ states is preserved, not affected by the sign change of $\lambda$. Hence the surface states still exist for Cu$_2$Se, as shown in Fig.~\ref{fig3}(b), although they are submerged in the bulk states. Importantly, the spin-momentum-locking pattern here becomes left-handed (Fig.~\ref{fig3}(c)), which is consistent with the positive SOC.

\begin{figure}[tbp]
\centerline{\includegraphics[clip,width=9.2cm]{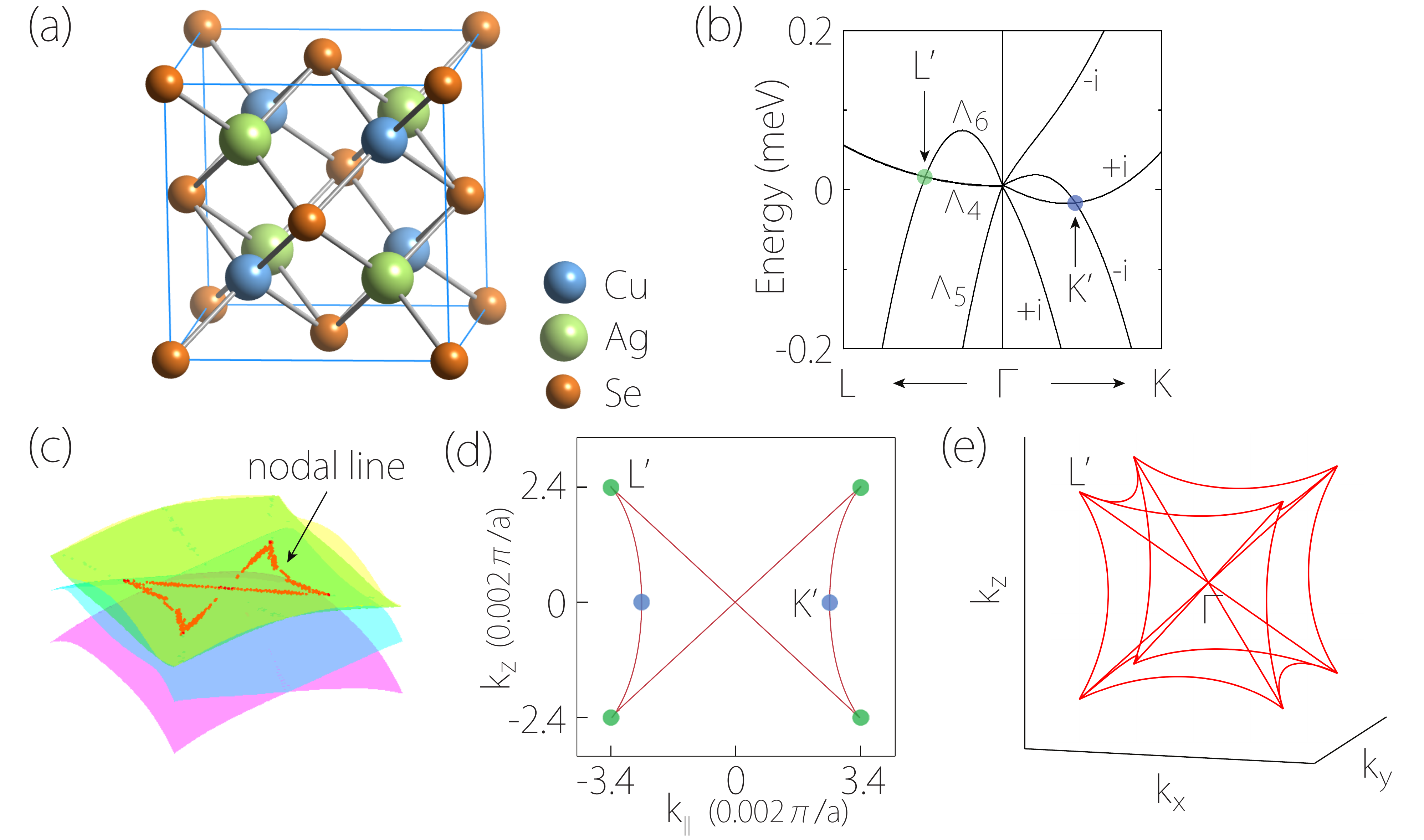}}
\caption{CuAgSe's (a) crystal structure, and (b) low-energy band structure around $\Gamma$ point. (c) Dispersion of the ($j=3/2$) low-energy bands in [110] mirror plane, showing (d) a butterfly-shaped nodal line, where $L'$ and $K'$ correspond to the crossing points marked in (b). (e) Combining all the butterfly nodal lines leads to a nodal box. }
\label{fig4}
\end{figure}

{\color{blue}{\em  Nodal-box and Weyl semimetals.}}---Breaking inversion symmetry would split the spin-degenerate bands with SOC, adding new ingredients to the physics. Consider CuAgSe in which the inversion symmetry is naturally broken (see Fig.~\ref{fig4}(a)). Similar to Cu$_2$Se, its $t_{2g}$ states have an effectively positive SOC, hence the system is also a semimetal, as shown in Fig.~\ref{fig4}(b). In Cu$_2$Se, the conduction band and the valence band touch at a single point, but we shall see that in CuAgSe, the splitting of the $j=3/2$ quartet near $\Gamma$-point leads to nodal-line band-crossings.

The symmetry group contains six mirror planes which may be collectively denoted as $M_{\{110\}}$. In Fig.~\ref{fig4}(b), one observes that the four states in the quartet fully split along the $\Gamma$-$K$ line, on which the two middle bands cross each other at a point $K'$ near the Fermi level. Meanwhile, along the $\Gamma$-$L$ direction, two middle bands become degenerate, which then cross the upper band at a triply-degenerate point (labeled as $L'$). A careful scan of the band structure around $\Gamma$-point shows that these crossing points are not isolated (see Fig.~\ref{fig4}(c)). Remarkably, in each $M_{\{110\}}$ mirror plane, the crossings near Fermi level form a butterfly-shaped nodal line, which is possible since it is formed by the pair-wise crossings of three bands. This is shown in Fig.~\ref{fig4}(d): along the diagonal (i.e. $\Gamma$-$L$) direction, the crossing is protected and pinned by the $C_{3v}$ symmetry such that the two states at each point on this line form a 2D irreducible representation $\Lambda_4$ ($E_{1/2}$); whereas for the arcs connecting the $L'$ points, they are protected by the mirror plane because the two crossing bands have opposite mirror eigenvalues $\pm i$ (see Fig.~\ref{fig4}(b)). Note that unlike most previously identified nodal lines which are unstable under SOC~\cite{PhysRevB.92.045108,PhysRevLett.115.036806,PhysRevLett.115.036807,Chen2015}, the butterfly nodal line here is robust against SOC, and actually it appears only when SOC is included. Combining the `butterflies' from all the mirror planes leads to the interesting nodal-box pattern shown in Fig.~\ref{fig4}(e).

\begin{figure}[tbp]
\centerline{\includegraphics[clip,width=9.2cm]{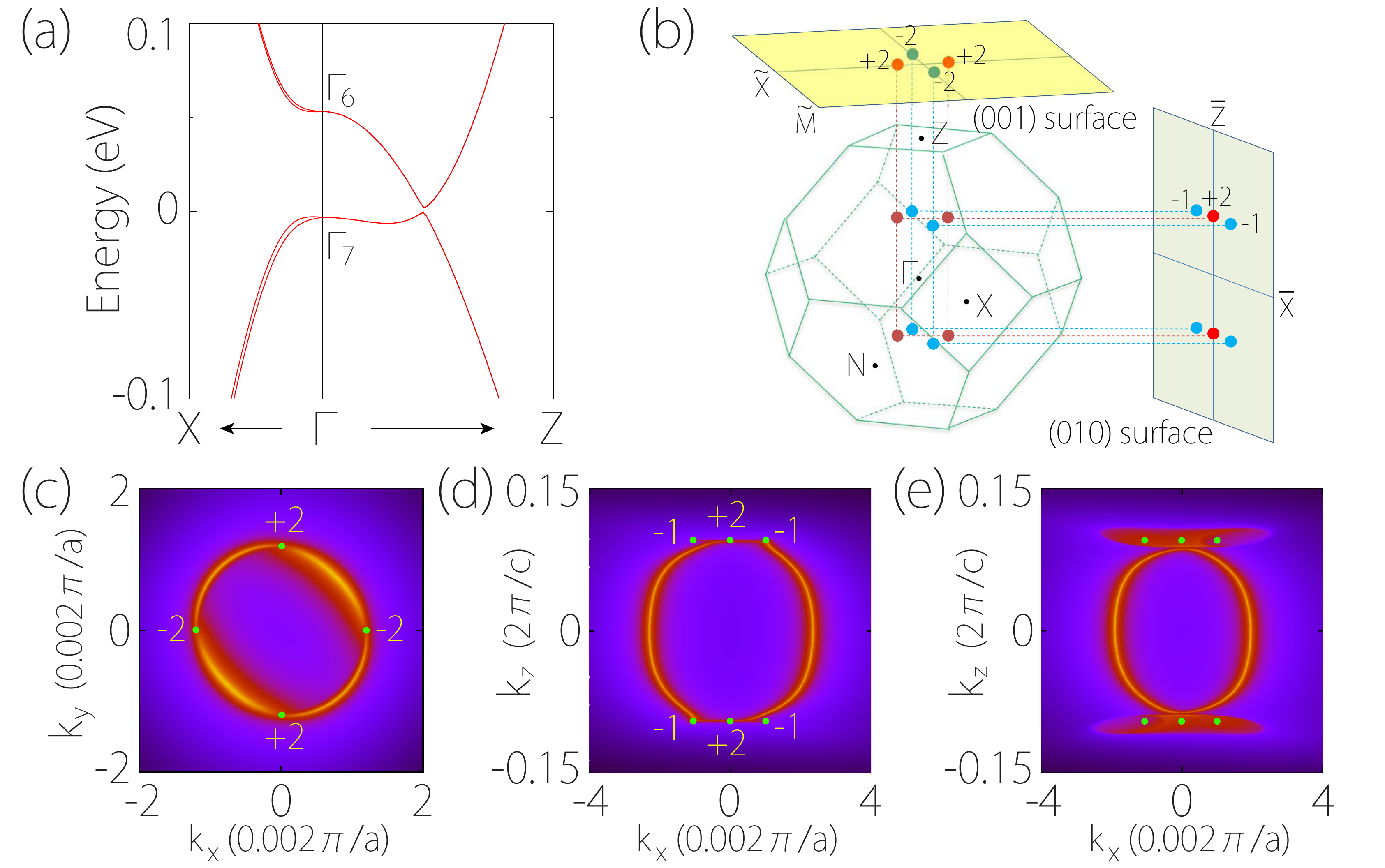}}
\caption{CuAgSe under 3\% uniaxial strain. (a) Band structure around $\Gamma$ point. A small gap is opened along $\Gamma$-$Z$. (b) Schematic view of 4 pairs of Weyl points in the Brillouin zone. (c) Fermi arcs on the (001) surface form a closed loop. (d) For (010) surface, Fermi arcs coexist with Dirac-cone surface states. (e) shows states at energy of $-5$ meV, where the Dirac-cone states can be cearly distinguished. The green dots mark the surface-projections of Weyl points.}
\label{fig5}
\end{figure}

It has been theoretically argued that a Weyl semimetal phase must occur during the transition of an inversion-asymmetric system from a TI phase to a normal insulator phase~\cite{Liu2014}. For CuAgSe, we find that it is a TI under a small uniaxial compression, and it becomes a normal insulator under a tensile strain. Hence there must exist a Weyl semimetal phase in-between. We carefully monitor the band structure change under uniaxial strains and find that the butterfly nodal lines quickly disappear upon applying a tensile strain. The reduction of the cubic symmetry leads to a dramatic change in the low-energy bands. Figure~\ref{fig5}(a) shows the band structure at 3\% strain, in which the gap almost closes along the $\Gamma$-$Z$ line.
We perform a scan of the possible band-crossing points using a dense $k$-mesh and reveal that there exist four pairs of Weyl points at ($\pm$0.001235, 0, $\pm$0.1012) and (0, $\pm$0.001235, $\pm$0.1012) in unit of reciprocal lattice vectors. As schematically shown in Fig.~\ref{fig5}(b), the points in each time reversal pair (i.e., at opposite $k$-points) have the same chirality, and the four pairs are further connected by the two remaining mirror planes $M_{(110)}$ and $M_{(1\bar{1}0)}$ (two points connected by mirror have opposite chiralities), so all the eight points are tied by symmetry and must locate at the Fermi level. Moreover, there is no other coexisting electron/hole pockets. Such kind of Weyl semimetals is referred to as ``ideal"~\cite{Ruan2016a,Ruan2016}, regarded as good platforms for studying the interesting Weyl physics.

The surface of a Weyl semimetal features Fermi arcs connecting the surface-projected Weyl points.
On (001) surface, there are four projected points, each having a chirality of $\pm 2$ (see Fig.~\ref{fig5}(b)), hence there must be two arcs connected to each point. This is verified by the surface spectrum in Fig.~\ref{fig5}(c), where the Fermi arcs are connected into a loop. Meanwhile, for (010) surface, there are six projected Weyl points, of which the middle ones on the $k_z$-axis have chirality $+2$ and other four have chirality $-1$ (Fig.~\ref{fig5}(b)). In the slice at Fermi energy (Fig.~\ref{fig5}(d)), one observes the nearly straight Fermi arcs connecting each $(-1,+2,-1)$ triplet. But surprisingly, there appears an additional Fermi circle, which looks like the TI Dirac-cone surface states. By checking the surface state variation at constant energies around Fermi level, one confirms that the circle is indeed separated from the Fermi arcs (see Fig.~\ref{fig5}(e)). To explain its existence, we note that $k_z=0$ time reversal invariant plane is fully gapped in the bulk hence carries a $Z_2$ invariant. Using the Wilson loop method~\cite{YuR_Wilson,PhysRevB.83.035108,WengHM2014review}, we find that its $Z_2=1$, indicating that there must be a pair of gapless states in edge spectrum of the $k_z=0$ plane.
This means that, in Fig.~\ref{fig5}(d) and~\ref{fig5}(e), there must exist a time reversal pair on the $k_z=0$ line. Therefore, the existence of the Dirac-cone is in fact required by the $Z_2$ invariant. In retrospect, one notes that planes such as $(110)$ and $(1\bar{1}0)$ also have nontrivial $Z_2$, but this does not lead to an additional Dirac-cone for $(001)$ surface because the pattern of Fermi arcs in Fig.~\ref{fig5}(c) already satisfies the requirement.

{\color{blue}{\em Discussion.}}---In this work, we have revealed an interesting family of topological materials in which the sign of SOC controls topological phase. On the negative SOC side, these materials are the first $d$-orbital TI discovered to date. We note that a few materials like TlN and HgS have been proposed as TIs with negative SOC~\cite{Sheng_TlN,Virot:2011fn}, but their low-energy bands are still dominated by $s$ and $p$ orbitals.

The sign of SOC determines the helicity of Dirac-cone surface states like in Fig.~\ref{fig2}(e) and~\ref{fig3}(c), which can be detected by spin-resolved ARPES experiment. The ability to control the helicity gives us additional freedom in utilizing these states for spintronics applications. Interestingly, it has been proposed that gapless interface states will emerge when two surfaces with opposite helicities are contacted~\cite{Takahashi2011}. The effect may be explored in our identified materials.

As mentioned, almost all the identified nodal lines are unstable under SOC. Certain kind of robust nodal lines requires nonsymmorphic symmetries~\cite{Fang2016}.
The butterfly nodal line discovered here is distinct in that: (i) it is formed under SOC; (ii) it does not require nonsymmorphic symmetries; and (iii) it is from the pair-wise crossings of three bands, not just two bands. Moreover, the multiple `butterflies' constitute a novel nodal box under tetragonal symmetry.

Finally, the coexistence of Fermi arcs and Dirac-cone surface states was proposed at an interface between a Weyl semimetal and a TI~\cite{Grushin2015}, and is also proposed in a model study~\cite{Lau2017}. Our work predicts the first material for its realization. The intriguing surface-state patterns can be directly probed via ARPES experiment. Furthermore, the interference between the Fermi arcs and the Dirac-cone may also lead to signatures in the quasiparticle interference pattern detectable via scanning tunneling microscopy~\cite{Lau2017}.


\newpage

\begin{widetext}
\section{Supplemental Material}
\renewcommand{\theequation}{S\arabic{equation}}
\setcounter{equation}{0}
\renewcommand{\thefigure}{S\arabic{figure}}
\setcounter{figure}{0}
\renewcommand{\thetable}{S\arabic{table}}
\setcounter{table}{0}

\subsection{Computational Methods}
\label{methods}
The results presented in the main text are obtained by first-principles calculations as implemented in Vienna \textit{ab initio} simulation package (VASP) \cite{VASP1,VASP2} with the projector augmented wave (PAW) method \cite{paw}. The generalized gradient approximation (GGA) with Perdew-Burke-Ernzerhof (PBE) \cite{GGAPBE} realization were adopted for the exchange-correlation potential. The plane-wave cutoff energy was taken as 500 eV. The Monkhorst-Pack $k$-point mesh~\cite{PhysRevB.13.5188} of size $10\times10\times 10$ was used for Brillouin zone sampling.  The crystal structures were optimized until the forces on the ions were less than 0.01 eV/\AA. As the transition metal $d$ orbitals may have
notable correlation effects, we also validate our results by GGA+U method\cite{PhysRevB.57.1505}. On-site Hubbard U parameters ranging from 1 eV to 7 eV were tested. We find that the key features are qualitatively the same as the GGA results, which also agree with previous studies~\cite{ZhangPH2014,Rasander2013_Cu2Se}. So in the main text, we focus on the GGA results. From the DFT results, we construct the maximally localized Wannier functions (MLWF)~\cite{Wannier90T} for Cu (Ag) $s$, $d$ and S (Se) $p$ orbitals, and effective model Hamiltonian for bulk and semi-infinite layer are built to investigate the surface states.

In this work, we focus on the antifluorite structure ($\alpha$-phase, space group $Fm\bar{3}m$ (No.~225)) of the Cu$_2$S-family materials. For the band structure calculations, we take the experimental lattice parameters $a=5.725$ \AA\;  for Cu$_2$S~\cite{Cu2S_5.725}, $a=5.787$ \AA\;  for Cu$_2$Se~\cite{Cu2Se_constant}, and $a=5.96$ \AA\; for CuAgS~\cite{CuAgS_constant}. For CuAgSe, since we also need to analyze its result under strain, we used its optimized lattice parameter $a=6.17$ \AA\; in the calculation.

\subsection{Band structures with modified Becke-Johnson potential}
Considering the possible underestimation of band gap by GGA, we further check
the band structure by the hybrid functional approach with modified Becke-Johnson (mBJ) potential~\cite{MBJ} as implemented in the WIEN2K package~\cite{WIEN2K}. We find that the band inversion features are maintained. The band inversion energy between $t_{2g}$ and Cu-$4s$ bands are of about 0.1 eV, 0.24 eV, 0.53 eV and 0.26 eV for Cu$_2$S, Cu$_2$Se, CuAgSe and CuAgS, respectively. The topological phases are still maintained. As shown in Fig.~\ref{fig:mbj}, Cu$_2$S is still a negative SOC TI, with direct (indirect) band gap of 65 meV (30 meV); CuAgS also keeps the negative SOC band
structure with direct (indirect) band gap of 71 meV (38 meV); Cu$_2$Se and CuAgSe
are still topological semimetals.

\begin{figure}[tbp]
\centerline{\includegraphics[clip,scale=0.55,angle=0]{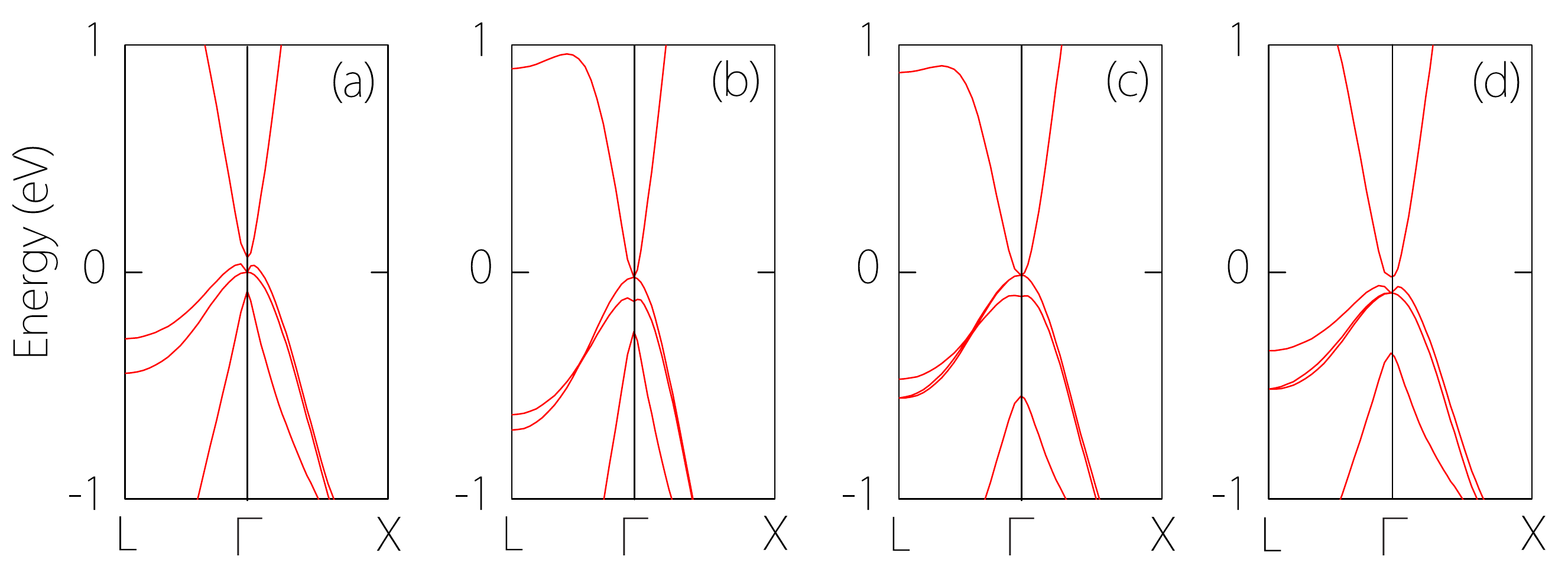}}
\caption{(Color online) Bulk band structures of the Cu$_2$S family materials using the modified Becke-Jonhson potential (mBJ) + SOC method for (a) Cu$_2$S, (b) Cu$_2$Se, (c) CuAgSe and (d) CuAgS.}
\label{fig:mbj}
\end{figure}

\subsection{$Z_2$ indices by Wilson loop method}
Band inversion is a necessary condition for a topological nontrivial band
structure, but it is not sufficient to identify a TI.
Topological invariant is a global character of the electronic structure
in the whole Brillouin zone. The definition of parity production for occupied bands at
TRIM points is a convenient method to distinguish TIs with inversion symmetry,
but it cannot be used for inversion-asymmetric systems such as CuAgSe.
The Wilson loop method can be employed for such cases. It traces the evolution of the Wannier
function centers~\cite{YuR_Wilson,PhysRevB.83.035108,WengHM2014review}. The Wannier center evolution for four representative planes (in the Brillouin zone) of CuAgSe are shown in Fig.~\ref{fig:wfc}, from which we can find the $Z_2$
indices are respectively 1 and 0 for $k_z=0$ plane and $k_z=\pi$ plane, indicating that there must be a pair of gapless states in edge spectrum of the $k_z=0$ plane. As discussed in the main text, this dictates the existence of a Dirac-cone 
on the $(010)$ surface. 
For the two mirror planes (110) and $(1\bar{1}0)$, we find $Z_2=1$. However, this does not lead to an additional Dirac-cone for $(001)$
surface because the pattern of the Fermi arcs already satisfies the requirement.

\begin{figure}[tbp]
\centerline{\includegraphics[clip,scale=0.55,angle=0]{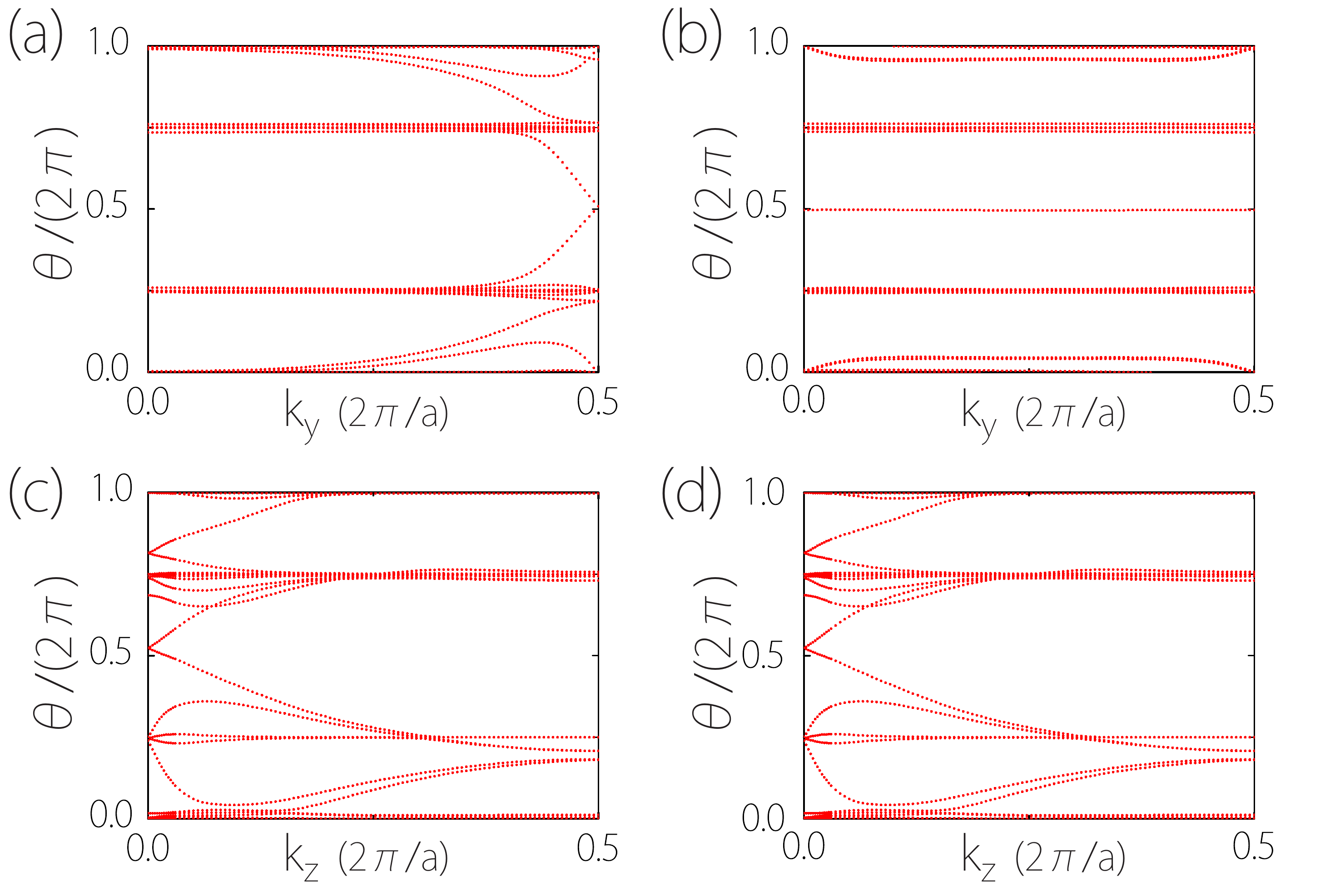}}
\caption{(Color online) The Wannier function center evolution for CuAgSe on the following planes in the Brillouin zone:
  (a) $k_z=0$, (b) $k_z=\pi$, (c) $(110)$ and (d) $(1\bar{1}0)$.}
\label{fig:wfc}
\end{figure}




\end{widetext}



\begin{thebibliography}{56}
\expandafter\ifx\csname natexlab\endcsname\relax\def\natexlab#1{#1}\fi
\expandafter\ifx\csname bibnamefont\endcsname\relax
  \def\bibnamefont#1{#1}\fi
\expandafter\ifx\csname bibfnamefont\endcsname\relax
  \def\bibfnamefont#1{#1}\fi
\expandafter\ifx\csname citenamefont\endcsname\relax
  \def\citenamefont#1{#1}\fi
\expandafter\ifx\csname url\endcsname\relax
  \def\url#1{\texttt{#1}}\fi
\expandafter\ifx\csname urlprefix\endcsname\relax\def\urlprefix{URL }\fi
\providecommand{\bibinfo}[2]{#2}
\providecommand{\eprint}[2][]{\url{#2}}

\bibitem[{\citenamefont{Hasan and Kane}(2010)}]{RevModPhys.82.3045}
\bibinfo{author}{\bibfnamefont{M.~Z.} \bibnamefont{Hasan}} \bibnamefont{and}
  \bibinfo{author}{\bibfnamefont{C.~L.} \bibnamefont{Kane}},
  \bibinfo{journal}{Rev. Mod. Phys.} \textbf{\bibinfo{volume}{82}},
  \bibinfo{pages}{3045} (\bibinfo{year}{2010}).

\bibitem[{\citenamefont{Qi and Zhang}(2011)}]{RevModPhys.83.1057}
\bibinfo{author}{\bibfnamefont{X.-L.} \bibnamefont{Qi}} \bibnamefont{and}
  \bibinfo{author}{\bibfnamefont{S.-C.} \bibnamefont{Zhang}},
  \bibinfo{journal}{Rev. Mod. Phys.} \textbf{\bibinfo{volume}{83}},
  \bibinfo{pages}{1057} (\bibinfo{year}{2011}).

\bibitem[{\citenamefont{Bansil et~al.}(2016)\citenamefont{Bansil, Lin, and
  Das}}]{RevModPhys.88.021004}
\bibinfo{author}{\bibfnamefont{A.}~\bibnamefont{Bansil}},
  \bibinfo{author}{\bibfnamefont{H.}~\bibnamefont{Lin}}, \bibnamefont{and}
  \bibinfo{author}{\bibfnamefont{T.}~\bibnamefont{Das}}, \bibinfo{journal}{Rev.
  Mod. Phys.} \textbf{\bibinfo{volume}{88}}, \bibinfo{pages}{021004}
  (\bibinfo{year}{2016}).

\bibitem[{\citenamefont{Volovik}(2003)}]{Volovik2003}
\bibinfo{author}{\bibfnamefont{G.~E.} \bibnamefont{Volovik}},
  \emph{\bibinfo{title}{The Universe in a Helium Droplet}}
  (\bibinfo{publisher}{Clarendon Press, Oxford}, \bibinfo{year}{2003}).

\bibitem[{\citenamefont{Zhao and Wang}(2013)}]{Zhao2013c}
\bibinfo{author}{\bibfnamefont{Y.~X.} \bibnamefont{Zhao}} \bibnamefont{and}
  \bibinfo{author}{\bibfnamefont{Z.~D.} \bibnamefont{Wang}},
  \bibinfo{journal}{Phys. Rev. Lett.} \textbf{\bibinfo{volume}{110}},
  \bibinfo{pages}{240404} (\bibinfo{year}{2013}).

\bibitem[{\citenamefont{Wan et~al.}(2011)\citenamefont{Wan, Turner, Vishwanath,
  and Savrasov}}]{WanXG_Weyl}
\bibinfo{author}{\bibfnamefont{X.}~\bibnamefont{Wan}},
  \bibinfo{author}{\bibfnamefont{A.~M.} \bibnamefont{Turner}},
  \bibinfo{author}{\bibfnamefont{A.}~\bibnamefont{Vishwanath}},
  \bibnamefont{and} \bibinfo{author}{\bibfnamefont{S.~Y.}
  \bibnamefont{Savrasov}}, \bibinfo{journal}{Phys. Rev. B}
  \textbf{\bibinfo{volume}{83}}, \bibinfo{pages}{205101}
  (\bibinfo{year}{2011}).

\bibitem[{\citenamefont{Murakami}(2007)}]{Murakami2007}
\bibinfo{author}{\bibfnamefont{S.}~\bibnamefont{Murakami}},
  \bibinfo{journal}{New J. Phys.} \textbf{\bibinfo{volume}{9}},
  \bibinfo{pages}{356} (\bibinfo{year}{2007}).

\bibitem[{\citenamefont{Weng et~al.}(2015{\natexlab{a}})\citenamefont{Weng,
  Fang, Fang, Bernevig, and Dai}}]{WengHM_2015TaAs}
\bibinfo{author}{\bibfnamefont{H.}~\bibnamefont{Weng}},
  \bibinfo{author}{\bibfnamefont{C.}~\bibnamefont{Fang}},
  \bibinfo{author}{\bibfnamefont{Z.}~\bibnamefont{Fang}},
  \bibinfo{author}{\bibfnamefont{B.~A.} \bibnamefont{Bernevig}},
  \bibnamefont{and} \bibinfo{author}{\bibfnamefont{X.}~\bibnamefont{Dai}},
  \bibinfo{journal}{Phys. Rev. X} \textbf{\bibinfo{volume}{5}},
  \bibinfo{pages}{011029} (\bibinfo{year}{2015}{\natexlab{a}}).

\bibitem[{\citenamefont{Huang et~al.}(2015)\citenamefont{Huang, Xu, Belopolski,
  Lee, Chang, Wang, Alidoust, Bian, Neupane, Zhang et~al.}}]{Huang2015}
\bibinfo{author}{\bibfnamefont{S.-M.} \bibnamefont{Huang}},
  \bibinfo{author}{\bibfnamefont{S.-Y.} \bibnamefont{Xu}},
  \bibinfo{author}{\bibfnamefont{I.}~\bibnamefont{Belopolski}},
  \bibinfo{author}{\bibfnamefont{C.-C.} \bibnamefont{Lee}},
  \bibinfo{author}{\bibfnamefont{G.}~\bibnamefont{Chang}},
  \bibinfo{author}{\bibfnamefont{B.}~\bibnamefont{Wang}},
  \bibinfo{author}{\bibfnamefont{N.}~\bibnamefont{Alidoust}},
  \bibinfo{author}{\bibfnamefont{G.}~\bibnamefont{Bian}},
  \bibinfo{author}{\bibfnamefont{M.}~\bibnamefont{Neupane}},
  \bibinfo{author}{\bibfnamefont{C.}~\bibnamefont{Zhang}},
  \bibnamefont{et~al.}, \bibinfo{journal}{Nat Commun}
  \textbf{\bibinfo{volume}{6}},  (\bibinfo{year}{2015}).

\bibitem[{\citenamefont{Lv et~al.}(2015)\citenamefont{Lv, Weng, Fu, Wang, Miao,
  Ma, Richard, Huang, Zhao, Chen et~al.}}]{Lv2015}
\bibinfo{author}{\bibfnamefont{B.~Q.} \bibnamefont{Lv}},
  \bibinfo{author}{\bibfnamefont{H.~M.} \bibnamefont{Weng}},
  \bibinfo{author}{\bibfnamefont{B.~B.} \bibnamefont{Fu}},
  \bibinfo{author}{\bibfnamefont{X.~P.} \bibnamefont{Wang}},
  \bibinfo{author}{\bibfnamefont{H.}~\bibnamefont{Miao}},
  \bibinfo{author}{\bibfnamefont{J.}~\bibnamefont{Ma}},
  \bibinfo{author}{\bibfnamefont{P.}~\bibnamefont{Richard}},
  \bibinfo{author}{\bibfnamefont{X.~C.} \bibnamefont{Huang}},
  \bibinfo{author}{\bibfnamefont{L.~X.} \bibnamefont{Zhao}},
  \bibinfo{author}{\bibfnamefont{G.~F.} \bibnamefont{Chen}},
  \bibnamefont{et~al.}, \bibinfo{journal}{Phys. Rev. X}
  \textbf{\bibinfo{volume}{5}}, \bibinfo{pages}{031013} (\bibinfo{year}{2015}).

\bibitem[{\citenamefont{Xu et~al.}(2015{\natexlab{a}})\citenamefont{Xu,
  Belopolski, Alidoust, Neupane, Bian, Zhang, Sankar, Chang, Yuan, Lee
  et~al.}}]{Xu613}
\bibinfo{author}{\bibfnamefont{S.-Y.} \bibnamefont{Xu}},
  \bibinfo{author}{\bibfnamefont{I.}~\bibnamefont{Belopolski}},
  \bibinfo{author}{\bibfnamefont{N.}~\bibnamefont{Alidoust}},
  \bibinfo{author}{\bibfnamefont{M.}~\bibnamefont{Neupane}},
  \bibinfo{author}{\bibfnamefont{G.}~\bibnamefont{Bian}},
  \bibinfo{author}{\bibfnamefont{C.}~\bibnamefont{Zhang}},
  \bibinfo{author}{\bibfnamefont{R.}~\bibnamefont{Sankar}},
  \bibinfo{author}{\bibfnamefont{G.}~\bibnamefont{Chang}},
  \bibinfo{author}{\bibfnamefont{Z.}~\bibnamefont{Yuan}},
  \bibinfo{author}{\bibfnamefont{C.-C.} \bibnamefont{Lee}},
  \bibnamefont{et~al.}, \bibinfo{journal}{Science}
  \textbf{\bibinfo{volume}{349}}, \bibinfo{pages}{613}
  (\bibinfo{year}{2015}{\natexlab{a}}).

\bibitem[{\citenamefont{Xu et~al.}(2015{\natexlab{b}})\citenamefont{Xu,
  Alidoust, Belopolski, Yuan, Bian, Chang, Zheng, Strocov, Sanchez, Chang
  et~al.}}]{Xu2015a}
\bibinfo{author}{\bibfnamefont{S.-Y.} \bibnamefont{Xu}},
  \bibinfo{author}{\bibfnamefont{N.}~\bibnamefont{Alidoust}},
  \bibinfo{author}{\bibfnamefont{I.}~\bibnamefont{Belopolski}},
  \bibinfo{author}{\bibfnamefont{Z.}~\bibnamefont{Yuan}},
  \bibinfo{author}{\bibfnamefont{G.}~\bibnamefont{Bian}},
  \bibinfo{author}{\bibfnamefont{T.-R.} \bibnamefont{Chang}},
  \bibinfo{author}{\bibfnamefont{H.}~\bibnamefont{Zheng}},
  \bibinfo{author}{\bibfnamefont{V.~N.} \bibnamefont{Strocov}},
  \bibinfo{author}{\bibfnamefont{D.~S.} \bibnamefont{Sanchez}},
  \bibinfo{author}{\bibfnamefont{G.}~\bibnamefont{Chang}},
  \bibnamefont{et~al.}, \bibinfo{journal}{Nat Phys}
  \textbf{\bibinfo{volume}{11}}, \bibinfo{pages}{748}
  (\bibinfo{year}{2015}{\natexlab{b}}).

\bibitem[{\citenamefont{Yang et~al.}(2015)\citenamefont{Yang, Liu, Sun, Peng,
  Yang, Zhang, Zhou, Zhang, Guo, Rahn et~al.}}]{Yang2015}
\bibinfo{author}{\bibfnamefont{L.~X.} \bibnamefont{Yang}},
  \bibinfo{author}{\bibfnamefont{Z.~K.} \bibnamefont{Liu}},
  \bibinfo{author}{\bibfnamefont{Y.}~\bibnamefont{Sun}},
  \bibinfo{author}{\bibfnamefont{H.}~\bibnamefont{Peng}},
  \bibinfo{author}{\bibfnamefont{H.~F.} \bibnamefont{Yang}},
  \bibinfo{author}{\bibfnamefont{T.}~\bibnamefont{Zhang}},
  \bibinfo{author}{\bibfnamefont{B.}~\bibnamefont{Zhou}},
  \bibinfo{author}{\bibfnamefont{Y.}~\bibnamefont{Zhang}},
  \bibinfo{author}{\bibfnamefont{Y.~F.} \bibnamefont{Guo}},
  \bibinfo{author}{\bibfnamefont{M.}~\bibnamefont{Rahn}}, \bibnamefont{et~al.},
  \bibinfo{journal}{Nat Phys} \textbf{\bibinfo{volume}{11}},
  \bibinfo{pages}{728} (\bibinfo{year}{2015}).

\bibitem[{\citenamefont{Weng et~al.}(2015{\natexlab{b}})\citenamefont{Weng,
  Liang, Xu, Yu, Fang, Dai, and Kawazoe}}]{PhysRevB.92.045108}
\bibinfo{author}{\bibfnamefont{H.}~\bibnamefont{Weng}},
  \bibinfo{author}{\bibfnamefont{Y.}~\bibnamefont{Liang}},
  \bibinfo{author}{\bibfnamefont{Q.}~\bibnamefont{Xu}},
  \bibinfo{author}{\bibfnamefont{R.}~\bibnamefont{Yu}},
  \bibinfo{author}{\bibfnamefont{Z.}~\bibnamefont{Fang}},
  \bibinfo{author}{\bibfnamefont{X.}~\bibnamefont{Dai}}, \bibnamefont{and}
  \bibinfo{author}{\bibfnamefont{Y.}~\bibnamefont{Kawazoe}},
  \bibinfo{journal}{Phys. Rev. B} \textbf{\bibinfo{volume}{92}},
  \bibinfo{pages}{045108} (\bibinfo{year}{2015}{\natexlab{b}}).

\bibitem[{\citenamefont{Kim et~al.}(2015)\citenamefont{Kim, Wieder, Kane, and
  Rappe}}]{PhysRevLett.115.036806}
\bibinfo{author}{\bibfnamefont{Y.}~\bibnamefont{Kim}},
  \bibinfo{author}{\bibfnamefont{B.~J.} \bibnamefont{Wieder}},
  \bibinfo{author}{\bibfnamefont{C.~L.} \bibnamefont{Kane}}, \bibnamefont{and}
  \bibinfo{author}{\bibfnamefont{A.~M.} \bibnamefont{Rappe}},
  \bibinfo{journal}{Phys. Rev. Lett.} \textbf{\bibinfo{volume}{115}},
  \bibinfo{pages}{036806} (\bibinfo{year}{2015}).

\bibitem[{\citenamefont{Yu et~al.}(2015)\citenamefont{Yu, Weng, Fang, Dai, and
  Hu}}]{PhysRevLett.115.036807}
\bibinfo{author}{\bibfnamefont{R.}~\bibnamefont{Yu}},
  \bibinfo{author}{\bibfnamefont{H.}~\bibnamefont{Weng}},
  \bibinfo{author}{\bibfnamefont{Z.}~\bibnamefont{Fang}},
  \bibinfo{author}{\bibfnamefont{X.}~\bibnamefont{Dai}}, \bibnamefont{and}
  \bibinfo{author}{\bibfnamefont{X.}~\bibnamefont{Hu}}, \bibinfo{journal}{Phys.
  Rev. Lett.} \textbf{\bibinfo{volume}{115}}, \bibinfo{pages}{036807}
  (\bibinfo{year}{2015}).

\bibitem[{\citenamefont{Chen et~al.}(2015)\citenamefont{Chen, Xie, Yang, Pan,
  Zhang, Cohen, and Zhang}}]{Chen2015}
\bibinfo{author}{\bibfnamefont{Y.}~\bibnamefont{Chen}},
  \bibinfo{author}{\bibfnamefont{Y.}~\bibnamefont{Xie}},
  \bibinfo{author}{\bibfnamefont{S.~A.} \bibnamefont{Yang}},
  \bibinfo{author}{\bibfnamefont{H.}~\bibnamefont{Pan}},
  \bibinfo{author}{\bibfnamefont{F.}~\bibnamefont{Zhang}},
  \bibinfo{author}{\bibfnamefont{M.~L.} \bibnamefont{Cohen}}, \bibnamefont{and}
  \bibinfo{author}{\bibfnamefont{S.}~\bibnamefont{Zhang}},
  \bibinfo{journal}{Nano Lett.} \textbf{\bibinfo{volume}{15}},
  \bibinfo{pages}{6974} (\bibinfo{year}{2015}).

\bibitem[{\citenamefont{Weng et~al.}(2016{\natexlab{a}})\citenamefont{Weng,
  Fang, Fang, and Dai}}]{PhysRevB.93.241202}
\bibinfo{author}{\bibfnamefont{H.}~\bibnamefont{Weng}},
  \bibinfo{author}{\bibfnamefont{C.}~\bibnamefont{Fang}},
  \bibinfo{author}{\bibfnamefont{Z.}~\bibnamefont{Fang}}, \bibnamefont{and}
  \bibinfo{author}{\bibfnamefont{X.}~\bibnamefont{Dai}},
  \bibinfo{journal}{Phys. Rev. B} \textbf{\bibinfo{volume}{93}},
  \bibinfo{pages}{241202} (\bibinfo{year}{2016}{\natexlab{a}}).

\bibitem[{\citenamefont{Weng et~al.}(2016{\natexlab{b}})\citenamefont{Weng,
  Fang, Fang, and Dai}}]{PhysRevB.94.165201}
\bibinfo{author}{\bibfnamefont{H.}~\bibnamefont{Weng}},
  \bibinfo{author}{\bibfnamefont{C.}~\bibnamefont{Fang}},
  \bibinfo{author}{\bibfnamefont{Z.}~\bibnamefont{Fang}}, \bibnamefont{and}
  \bibinfo{author}{\bibfnamefont{X.}~\bibnamefont{Dai}},
  \bibinfo{journal}{Phys. Rev. B} \textbf{\bibinfo{volume}{94}},
  \bibinfo{pages}{165201} (\bibinfo{year}{2016}{\natexlab{b}}).

\bibitem[{\citenamefont{Zhu et~al.}(2016)\citenamefont{Zhu, Winkler, Wu, Li,
  and Soluyanov}}]{PhysRevX.6.031003}
\bibinfo{author}{\bibfnamefont{Z.}~\bibnamefont{Zhu}},
  \bibinfo{author}{\bibfnamefont{G.~W.} \bibnamefont{Winkler}},
  \bibinfo{author}{\bibfnamefont{Q.}~\bibnamefont{Wu}},
  \bibinfo{author}{\bibfnamefont{J.}~\bibnamefont{Li}}, \bibnamefont{and}
  \bibinfo{author}{\bibfnamefont{A.~A.} \bibnamefont{Soluyanov}},
  \bibinfo{journal}{Phys. Rev. X} \textbf{\bibinfo{volume}{6}},
  \bibinfo{pages}{031003} (\bibinfo{year}{2016}).

\bibitem[{\citenamefont{Feng and Yao}(2012)}]{YaoYG2012_TIreview}
\bibinfo{author}{\bibfnamefont{W.}~\bibnamefont{Feng}} \bibnamefont{and}
  \bibinfo{author}{\bibfnamefont{Y.}~\bibnamefont{Yao}}, \bibinfo{journal}{Sci.
  China Phys. Mech. Astron.} \textbf{\bibinfo{volume}{55}},
  \bibinfo{pages}{2199} (\bibinfo{year}{2012}).

\bibitem[{\citenamefont{Ando}(2013)}]{Ando2013}
\bibinfo{author}{\bibfnamefont{Y.}~\bibnamefont{Ando}}, \bibinfo{journal}{J.
  Phys. Soc. Jpn.} \textbf{\bibinfo{volume}{82}} (\bibinfo{year}{2013}).

\bibitem[{\citenamefont{Vidal et~al.}(2012)\citenamefont{Vidal, Zhang,
  Stevanovi{\'c}, Luo, and Zunger}}]{Vidal:2012cr}
\bibinfo{author}{\bibfnamefont{J.}~\bibnamefont{Vidal}},
  \bibinfo{author}{\bibfnamefont{X.}~\bibnamefont{Zhang}},
  \bibinfo{author}{\bibfnamefont{V.}~\bibnamefont{Stevanovi{\'c}}},
  \bibinfo{author}{\bibfnamefont{J.-W.} \bibnamefont{Luo}}, \bibnamefont{and}
  \bibinfo{author}{\bibfnamefont{A.}~\bibnamefont{Zunger}},
  \bibinfo{journal}{Phys. Rev. B} \textbf{\bibinfo{volume}{86}},
  \bibinfo{pages}{075316} (\bibinfo{year}{2012}).

\bibitem[{\citenamefont{Sheng et~al.}(2014)\citenamefont{Sheng, Wang, Yu, Weng,
  Fang, and Dai}}]{Sheng_TlN}
\bibinfo{author}{\bibfnamefont{X.-L.} \bibnamefont{Sheng}},
  \bibinfo{author}{\bibfnamefont{Z.}~\bibnamefont{Wang}},
  \bibinfo{author}{\bibfnamefont{R.}~\bibnamefont{Yu}},
  \bibinfo{author}{\bibfnamefont{H.}~\bibnamefont{Weng}},
  \bibinfo{author}{\bibfnamefont{Z.}~\bibnamefont{Fang}}, \bibnamefont{and}
  \bibinfo{author}{\bibfnamefont{X.}~\bibnamefont{Dai}},
  \bibinfo{journal}{Phys. Rev. B} \textbf{\bibinfo{volume}{90}},
  \bibinfo{pages}{245308} (\bibinfo{year}{2014}).

\bibitem[{\citenamefont{Virot et~al.}(2011)\citenamefont{Virot, Hayn, Richter,
  and van~den Brink}}]{Virot:2011fn}
\bibinfo{author}{\bibfnamefont{F.}~\bibnamefont{Virot}},
  \bibinfo{author}{\bibfnamefont{R.}~\bibnamefont{Hayn}},
  \bibinfo{author}{\bibfnamefont{M.}~\bibnamefont{Richter}}, \bibnamefont{and}
  \bibinfo{author}{\bibfnamefont{J.}~\bibnamefont{van~den Brink}},
  \bibinfo{journal}{Phys. Rev. Lett.} \textbf{\bibinfo{volume}{106}},
  \bibinfo{pages}{236806} (\bibinfo{year}{2011}).

\bibitem[{\citenamefont{Djurle}(1958)}]{Cu2S_5.725}
\bibinfo{author}{\bibfnamefont{S.}~\bibnamefont{Djurle}},
  \bibinfo{journal}{Acta Chem. Scand.} \textbf{\bibinfo{volume}{12}},
  \bibinfo{pages}{1415} (\bibinfo{year}{1958}).

\bibitem[{\citenamefont{Yamamoto and Kashida}(1991)}]{Cu2Se_constant}
\bibinfo{author}{\bibfnamefont{K.}~\bibnamefont{Yamamoto}} \bibnamefont{and}
  \bibinfo{author}{\bibfnamefont{S.}~\bibnamefont{Kashida}},
  \bibinfo{journal}{J. Solid State Chem.} \textbf{\bibinfo{volume}{93}},
  \bibinfo{pages}{202 } (\bibinfo{year}{1991}).

\bibitem[{\citenamefont{Trots et~al.}(2007)\citenamefont{Trots, Senyshyn,
  Mikhailova, Knapp, Baehtz, Hoelzel, and Fuess}}]{CuAgS_constant}
\bibinfo{author}{\bibfnamefont{D.~M.} \bibnamefont{Trots}},
  \bibinfo{author}{\bibfnamefont{A.}~\bibnamefont{Senyshyn}},
  \bibinfo{author}{\bibfnamefont{D.~A.} \bibnamefont{Mikhailova}},
  \bibinfo{author}{\bibfnamefont{M.}~\bibnamefont{Knapp}},
  \bibinfo{author}{\bibfnamefont{C.}~\bibnamefont{Baehtz}},
  \bibinfo{author}{\bibfnamefont{M.}~\bibnamefont{Hoelzel}}, \bibnamefont{and}
  \bibinfo{author}{\bibfnamefont{H.}~\bibnamefont{Fuess}}, \bibinfo{journal}{J.
  Phys. Condens. Matter} \textbf{\bibinfo{volume}{19}}, \bibinfo{pages}{136204}
  (\bibinfo{year}{2007}).

\bibitem[{SI()}]{SI}
\bibinfo{note}{See Supplemental Material.}

\bibitem[{\citenamefont{Cardona}(1963)}]{Cardona1963}
\bibinfo{author}{\bibfnamefont{M.}~\bibnamefont{Cardona}},
  \bibinfo{journal}{Phys. Rev.} \textbf{\bibinfo{volume}{129}},
  \bibinfo{pages}{69} (\bibinfo{year}{1963}).

\bibitem[{\citenamefont{Shindo et~al.}(1965)\citenamefont{Shindo, Morita, and
  Kamimura}}]{Shindo1965}
\bibinfo{author}{\bibfnamefont{K.}~\bibnamefont{Shindo}},
  \bibinfo{author}{\bibfnamefont{A.}~\bibnamefont{Morita}}, \bibnamefont{and}
  \bibinfo{author}{\bibfnamefont{H.}~\bibnamefont{Kamimura}},
  \bibinfo{journal}{J. Phys. Soc. Jpn.} \textbf{\bibinfo{volume}{20}},
  \bibinfo{pages}{2054} (\bibinfo{year}{1965}).

\bibitem[{\citenamefont{Fu and Kane}(2007)}]{Fu2007}
\bibinfo{author}{\bibfnamefont{L.}~\bibnamefont{Fu}} \bibnamefont{and}
  \bibinfo{author}{\bibfnamefont{C.~L.} \bibnamefont{Kane}},
  \bibinfo{journal}{Phys. Rev. B} \textbf{\bibinfo{volume}{76}},
  \bibinfo{pages}{045302} (\bibinfo{year}{2007}).

\bibitem[{\citenamefont{Zhang et~al.}(2010)\citenamefont{Zhang, Yu, Zhang, Dai,
  and Fang}}]{ZhangW_NJP}
\bibinfo{author}{\bibfnamefont{W.}~\bibnamefont{Zhang}},
  \bibinfo{author}{\bibfnamefont{R.}~\bibnamefont{Yu}},
  \bibinfo{author}{\bibfnamefont{H.-J.} \bibnamefont{Zhang}},
  \bibinfo{author}{\bibfnamefont{X.}~\bibnamefont{Dai}}, \bibnamefont{and}
  \bibinfo{author}{\bibfnamefont{Z.}~\bibnamefont{Fang}}, \bibinfo{journal}{New
  J. Phys.} \textbf{\bibinfo{volume}{12}}, \bibinfo{pages}{065013}
  (\bibinfo{year}{2010}).

\bibitem[{\citenamefont{Zhang et~al.}(2013)\citenamefont{Zhang, Liu, and
  Zhang}}]{PhysRevLett.111.066801}
\bibinfo{author}{\bibfnamefont{H.}~\bibnamefont{Zhang}},
  \bibinfo{author}{\bibfnamefont{C.-X.} \bibnamefont{Liu}}, \bibnamefont{and}
  \bibinfo{author}{\bibfnamefont{S.-C.} \bibnamefont{Zhang}},
  \bibinfo{journal}{Phys. Rev. Lett.} \textbf{\bibinfo{volume}{111}},
  \bibinfo{pages}{066801} (\bibinfo{year}{2013}).

\bibitem[{\citenamefont{Liu and Vanderbilt}(2014)}]{Liu2014}
\bibinfo{author}{\bibfnamefont{J.}~\bibnamefont{Liu}} \bibnamefont{and}
  \bibinfo{author}{\bibfnamefont{D.}~\bibnamefont{Vanderbilt}},
  \bibinfo{journal}{Phys. Rev. B} \textbf{\bibinfo{volume}{90}},
  \bibinfo{pages}{155316} (\bibinfo{year}{2014}).

\bibitem[{\citenamefont{Ruan et~al.}(2016{\natexlab{a}})\citenamefont{Ruan,
  Jian, Yao, Zhang, Zhang, and Xing}}]{Ruan2016a}
\bibinfo{author}{\bibfnamefont{J.}~\bibnamefont{Ruan}},
  \bibinfo{author}{\bibfnamefont{S.-K.} \bibnamefont{Jian}},
  \bibinfo{author}{\bibfnamefont{H.}~\bibnamefont{Yao}},
  \bibinfo{author}{\bibfnamefont{H.}~\bibnamefont{Zhang}},
  \bibinfo{author}{\bibfnamefont{S.-C.} \bibnamefont{Zhang}}, \bibnamefont{and}
  \bibinfo{author}{\bibfnamefont{D.}~\bibnamefont{Xing}},
  \bibinfo{journal}{Nat. Commun.} \textbf{\bibinfo{volume}{7}},
  \bibinfo{pages}{11136} (\bibinfo{year}{2016}{\natexlab{a}}),
  \eprint{1511.08284}.

\bibitem[{\citenamefont{Ruan et~al.}(2016{\natexlab{b}})\citenamefont{Ruan,
  Jian, Zhang, Yao, Zhang, Zhang, and Xing}}]{Ruan2016}
\bibinfo{author}{\bibfnamefont{J.}~\bibnamefont{Ruan}},
  \bibinfo{author}{\bibfnamefont{S.-K.} \bibnamefont{Jian}},
  \bibinfo{author}{\bibfnamefont{D.}~\bibnamefont{Zhang}},
  \bibinfo{author}{\bibfnamefont{H.}~\bibnamefont{Yao}},
  \bibinfo{author}{\bibfnamefont{H.}~\bibnamefont{Zhang}},
  \bibinfo{author}{\bibfnamefont{S.-C.} \bibnamefont{Zhang}}, \bibnamefont{and}
  \bibinfo{author}{\bibfnamefont{D.}~\bibnamefont{Xing}},
  \bibinfo{journal}{Phys. Rev. Lett.} \textbf{\bibinfo{volume}{116}},
  \bibinfo{pages}{226801} (\bibinfo{year}{2016}{\natexlab{b}}).

\bibitem[{\citenamefont{Yu et~al.}(2011)\citenamefont{Yu, Qi, Bernevig, Fang,
  and Dai}}]{YuR_Wilson}
\bibinfo{author}{\bibfnamefont{R.}~\bibnamefont{Yu}},
  \bibinfo{author}{\bibfnamefont{X.~L.} \bibnamefont{Qi}},
  \bibinfo{author}{\bibfnamefont{A.}~\bibnamefont{Bernevig}},
  \bibinfo{author}{\bibfnamefont{Z.}~\bibnamefont{Fang}}, \bibnamefont{and}
  \bibinfo{author}{\bibfnamefont{X.}~\bibnamefont{Dai}},
  \bibinfo{journal}{Phys. Rev. B} \textbf{\bibinfo{volume}{84}},
  \bibinfo{pages}{075119} (\bibinfo{year}{2011}).

\bibitem[{\citenamefont{Soluyanov and Vanderbilt}(2011)}]{PhysRevB.83.035108}
\bibinfo{author}{\bibfnamefont{A.~A.} \bibnamefont{Soluyanov}}
  \bibnamefont{and}
  \bibinfo{author}{\bibfnamefont{D.}~\bibnamefont{Vanderbilt}},
  \bibinfo{journal}{Phys. Rev. B} \textbf{\bibinfo{volume}{83}},
  \bibinfo{pages}{035108} (\bibinfo{year}{2011}).

\bibitem[{\citenamefont{Weng et~al.}(2014)\citenamefont{Weng, Dai, and
  Fang}}]{WengHM2014review}
\bibinfo{author}{\bibfnamefont{H.}~\bibnamefont{Weng}},
  \bibinfo{author}{\bibfnamefont{X.}~\bibnamefont{Dai}}, \bibnamefont{and}
  \bibinfo{author}{\bibfnamefont{Z.}~\bibnamefont{Fang}}, \bibinfo{journal}{MRS
  Bull.} \textbf{\bibinfo{volume}{39}}, \bibinfo{pages}{849}
  (\bibinfo{year}{2014}).

\bibitem[{\citenamefont{Takahashi and Murakami}(2011)}]{Takahashi2011}
\bibinfo{author}{\bibfnamefont{R.}~\bibnamefont{Takahashi}} \bibnamefont{and}
  \bibinfo{author}{\bibfnamefont{S.}~\bibnamefont{Murakami}},
  \bibinfo{journal}{Phys. Rev. Lett.} \textbf{\bibinfo{volume}{107}},
  \bibinfo{pages}{166805} (\bibinfo{year}{2011}).

\bibitem[{\citenamefont{Fang et~al.}(2016)\citenamefont{Fang, Weng, Dai, and
  Fang}}]{Fang2016}
\bibinfo{author}{\bibfnamefont{C.}~\bibnamefont{Fang}},
  \bibinfo{author}{\bibfnamefont{H.}~\bibnamefont{Weng}},
  \bibinfo{author}{\bibfnamefont{X.}~\bibnamefont{Dai}}, \bibnamefont{and}
  \bibinfo{author}{\bibfnamefont{Z.}~\bibnamefont{Fang}},
  \bibinfo{journal}{Chin. Phys. B} \textbf{\bibinfo{volume}{25}},
  \bibinfo{pages}{117106} (\bibinfo{year}{2016}).

\bibitem[{\citenamefont{Grushin et~al.}(2015)\citenamefont{Grushin, Venderbos,
  and Bardarson}}]{Grushin2015}
\bibinfo{author}{\bibfnamefont{A.~G.} \bibnamefont{Grushin}},
  \bibinfo{author}{\bibfnamefont{J.~W.~F.} \bibnamefont{Venderbos}},
  \bibnamefont{and} \bibinfo{author}{\bibfnamefont{J.~H.}
  \bibnamefont{Bardarson}}, \bibinfo{journal}{Phys. Rev. B}
  \textbf{\bibinfo{volume}{91}}, \bibinfo{pages}{121109}
  (\bibinfo{year}{2015}).

\bibitem[{Lau()}]{Lau2017}
\bibinfo{note}{A. Lau, J. van den Brink, and C. Ortix, arXiv:1701.01660.}

\expandafter\ifx\csname natexlab\endcsname\relax\def\natexlab#1{#1}\fi
\expandafter\ifx\csname bibnamefont\endcsname\relax
  \def\bibnamefont#1{#1}\fi
\expandafter\ifx\csname bibfnamefont\endcsname\relax
  \def\bibfnamefont#1{#1}\fi
\expandafter\ifx\csname citenamefont\endcsname\relax
  \def\citenamefont#1{#1}\fi
\expandafter\ifx\csname url\endcsname\relax
  \def\url#1{\texttt{#1}}\fi
\expandafter\ifx\csname urlprefix\endcsname\relax\def\urlprefix{URL }\fi
\providecommand{\bibinfo}[2]{#2}
\providecommand{\eprint}[2][]{\url{#2}}

\bibitem[{\citenamefont{Kresse and Furthm\"uller}(1996)}]{VASP1}
\bibinfo{author}{\bibfnamefont{G.}~\bibnamefont{Kresse}} \bibnamefont{and}
  \bibinfo{author}{\bibfnamefont{J.}~\bibnamefont{Furthm\"uller}},
  \bibinfo{journal}{Phys. Rev. B} \textbf{\bibinfo{volume}{54}},
  \bibinfo{pages}{11169} (\bibinfo{year}{1996}).

\bibitem[{\citenamefont{Kresse and Joubert}(1999)}]{VASP2}
\bibinfo{author}{\bibfnamefont{G.}~\bibnamefont{Kresse}} \bibnamefont{and}
  \bibinfo{author}{\bibfnamefont{D.}~\bibnamefont{Joubert}},
  \bibinfo{journal}{Phys. Rev. B} \textbf{\bibinfo{volume}{59}},
  \bibinfo{pages}{1758} (\bibinfo{year}{1999}).

\bibitem[{\citenamefont{Bl\"ochl}(1994)}]{paw}
\bibinfo{author}{\bibfnamefont{P.~E.} \bibnamefont{Bl\"ochl}},
  \bibinfo{journal}{Phys. Rev. B} \textbf{\bibinfo{volume}{50}},
  \bibinfo{pages}{17953} (\bibinfo{year}{1994}).

\bibitem[{\citenamefont{Perdew et~al.}(1996)\citenamefont{Perdew, Burke, and
  Ernzerhof}}]{GGAPBE}
\bibinfo{author}{\bibfnamefont{J.~P.} \bibnamefont{Perdew}},
  \bibinfo{author}{\bibfnamefont{K.}~\bibnamefont{Burke}}, \bibnamefont{and}
  \bibinfo{author}{\bibfnamefont{M.}~\bibnamefont{Ernzerhof}},
  \bibinfo{journal}{Phys. Rev. Lett.} \textbf{\bibinfo{volume}{77}},
  \bibinfo{pages}{3865} (\bibinfo{year}{1996}).

\bibitem[{\citenamefont{Monkhorst and Pack}(1976)}]{PhysRevB.13.5188}
\bibinfo{author}{\bibfnamefont{H.~J.} \bibnamefont{Monkhorst}}
  \bibnamefont{and} \bibinfo{author}{\bibfnamefont{J.~D.} \bibnamefont{Pack}},
  \bibinfo{journal}{Phys. Rev. B} \textbf{\bibinfo{volume}{13}},
  \bibinfo{pages}{5188} (\bibinfo{year}{1976}).

\bibitem[{\citenamefont{Dudarev et~al.}(1998)\citenamefont{Dudarev, Botton,
  Savrasov, Humphreys, and Sutton}}]{PhysRevB.57.1505}
\bibinfo{author}{\bibfnamefont{S.~L.} \bibnamefont{Dudarev}},
  \bibinfo{author}{\bibfnamefont{G.~A.} \bibnamefont{Botton}},
  \bibinfo{author}{\bibfnamefont{S.~Y.} \bibnamefont{Savrasov}},
  \bibinfo{author}{\bibfnamefont{C.~J.} \bibnamefont{Humphreys}},
  \bibnamefont{and} \bibinfo{author}{\bibfnamefont{A.~P.}
  \bibnamefont{Sutton}}, \bibinfo{journal}{Phys. Rev. B}
  \textbf{\bibinfo{volume}{57}}, \bibinfo{pages}{1505} (\bibinfo{year}{1998}).

\bibitem[{\citenamefont{Zhang et~al.}(2014)\citenamefont{Zhang, Wang, Xi, Qiu,
  Shi, Zhang, and Zhang}}]{ZhangPH2014}
\bibinfo{author}{\bibfnamefont{Y.}~\bibnamefont{Zhang}},
  \bibinfo{author}{\bibfnamefont{Y.}~\bibnamefont{Wang}},
  \bibinfo{author}{\bibfnamefont{L.}~\bibnamefont{Xi}},
  \bibinfo{author}{\bibfnamefont{R.}~\bibnamefont{Qiu}},
  \bibinfo{author}{\bibfnamefont{X.}~\bibnamefont{Shi}},
  \bibinfo{author}{\bibfnamefont{P.}~\bibnamefont{Zhang}}, \bibnamefont{and}
  \bibinfo{author}{\bibfnamefont{W.}~\bibnamefont{Zhang}}, \bibinfo{journal}{J.
  Chem. Phys.} \textbf{\bibinfo{volume}{140}} (\bibinfo{year}{2014}).

\bibitem[{\citenamefont{R{\aa}sander et~al.}(2013)\citenamefont{R{\aa}sander,
  Bergqvist, and Delin}}]{Rasander2013_Cu2Se}
\bibinfo{author}{\bibfnamefont{M.}~\bibnamefont{R{\aa}sander}},
  \bibinfo{author}{\bibfnamefont{L.}~\bibnamefont{Bergqvist}},
  \bibnamefont{and} \bibinfo{author}{\bibfnamefont{A.}~\bibnamefont{Delin}},
  \bibinfo{journal}{J. Phys. Condens. Matter} \textbf{\bibinfo{volume}{25}},
  \bibinfo{pages}{125503} (\bibinfo{year}{2013}).

\bibitem[{\citenamefont{Mostofi et~al.}(2014)\citenamefont{Mostofi, Yates,
  Pizzi, Lee, Souza, Vanderbilt, and Marzari}}]{Wannier90T}
\bibinfo{author}{\bibfnamefont{A.~A.} \bibnamefont{Mostofi}},
  \bibinfo{author}{\bibfnamefont{J.~R.} \bibnamefont{Yates}},
  \bibinfo{author}{\bibfnamefont{G.}~\bibnamefont{Pizzi}},
  \bibinfo{author}{\bibfnamefont{Y.-S.} \bibnamefont{Lee}},
  \bibinfo{author}{\bibfnamefont{I.}~\bibnamefont{Souza}},
  \bibinfo{author}{\bibfnamefont{D.}~\bibnamefont{Vanderbilt}},
  \bibnamefont{and} \bibinfo{author}{\bibfnamefont{N.}~\bibnamefont{Marzari}},
  \bibinfo{journal}{Comput. Phys. Comm.} \textbf{\bibinfo{volume}{185}},
  \bibinfo{pages}{2309 } (\bibinfo{year}{2014}).

\bibitem[{\citenamefont{Becke and Johnson}(2006)}]{MBJ}
\bibinfo{author}{\bibfnamefont{A.~D.} \bibnamefont{Becke}} \bibnamefont{and}
  \bibinfo{author}{\bibfnamefont{E.~R.} \bibnamefont{Johnson}},
  \bibinfo{journal}{J. Chem. Phys.} \textbf{\bibinfo{volume}{124}},
  \bibinfo{pages}{221101} (\bibinfo{year}{2006}).

\bibitem[{\citenamefont{Blaha et~al.}(2001)\citenamefont{Blaha, Schwarz,
  Madsen, D., and J.}}]{WIEN2K}
\bibinfo{author}{\bibfnamefont{P.}~\bibnamefont{Blaha}},
  \bibinfo{author}{\bibfnamefont{K.}~\bibnamefont{Schwarz}},
  \bibinfo{author}{\bibfnamefont{G.}~\bibnamefont{Madsen}},
  \bibinfo{author}{\bibfnamefont{K.}~\bibnamefont{D.}}, \bibnamefont{and}
  \bibinfo{author}{\bibfnamefont{L.}~\bibnamefont{J.}},
  \emph{\bibinfo{title}{WIEN2k, An Augmented Plane Wave + Local Orbitals
  Program for Calculating Crystal Properties}} (\bibinfo{year}{2001}), ISBN
  \bibinfo{isbn}{3-9501031-1-2}.


\end{thebibliography}



\end{document}